\documentclass[sigconf,nonacm]{acmart}
\AtBeginDocument{%
  }

\newcommand{\ie}{\textit{i.e.}}
\newcommand{\eg}{\textit{e.g.}}

\newcommand{\sysName}{\textsc{DOMSteer}}

\usepackage[dvipsnames]{xcolor}

\usepackage{duckuments}
\usepackage{enumitem}
\usepackage{multirow}

\usepackage{flushend}
\usepackage{listings}
\begin{document}


\title[Beyond Chat and Clicks: GUI Agents for In-Situ Assistance via Live Interface Transformation]{Beyond Chat and Clicks: GUI Agents for In-Situ Assistance via Live Interface Transformation}

\author{Pan Hao}
\email{pan00342@umn.edu}
\orcid{0009-0006-4473-5764}
\affiliation{%
  \institution{University of Minnesota}
  \city{Minneapolis}
  \state{MN}
  \country{USA}
}

\author{Rishi Selvakumaran}
\authornote{Both authors contributed equally to this work.}
\email{selva053@umn.edu}
\affiliation{%
  \institution{University of Minnesota}
  \city{Minneapolis}
  \state{MN}
  \country{USA}
}

\author{Jacob Sun}
\authornotemark[1]
\email{sun00572@umn.edu}
\affiliation{%
  \institution{University of Minnesota}
  \city{Minneapolis}
  \state{MN}
  \country{USA}
}

\author{Qianwen Wang}
\email{qianwen@umn.edu}
\orcid{0000-0002-1825-0097}
\affiliation{%
  \institution{University of Minnesota}
  \city{Minneapolis}
  \state{MN}
  \country{USA}
}

\renewcommand{\shortauthors}{Hao et al.}
\begin{abstract}
Complex visual interfaces are powerful yet have a steep learning curve, as users must navigate feature-rich visual interfaces while reasoning about domain-specific operations.  
Existing approaches either deliver assistance through a separate chat-based interaction, or require substantial application-specific engineering to build support natively into each interface. 
To address the gaps, we propose \textit{in-situ assistance}: a mode of support delivered directly within any live web interface through lightweight, browser-level interventions on the Document Object Model (DOM), without rebuilding the application or modifying its underlying logic.
We contribute a design space and a computational pipeline for DOM-mediated in-situ assistance, characterizing how GUI agents can insert, mutate, or recompose web elements to make the interface easier for users to understand, use, and navigate.
We instantiate \textit{in-situ assistance} in \sysName{}, a Chrome extension that interprets a user’s help request and live interface context, grounds it to relevant UI elements, and executes reversible DOM manipulations directly on the live page to deliver assistance, including contextual tooltips, control highlighting, layout reorganization. 
Quantitative evaluations on two complex visual interfaces show that \sysName{} delivers reliable and efficient in-situ assistance. 
Use cases and a comparative user study with  ChatGPTAtlas demonstrate the usability and effectiveness of \sysName{}.
Altogether, these findings point to a broader role for GUI agents: not just assisting from the sidelines, but actively reconfiguring live interfaces to support users in the moment.
\end{abstract}

\begin{teaserfigure}
  \includegraphics[width=\textwidth]{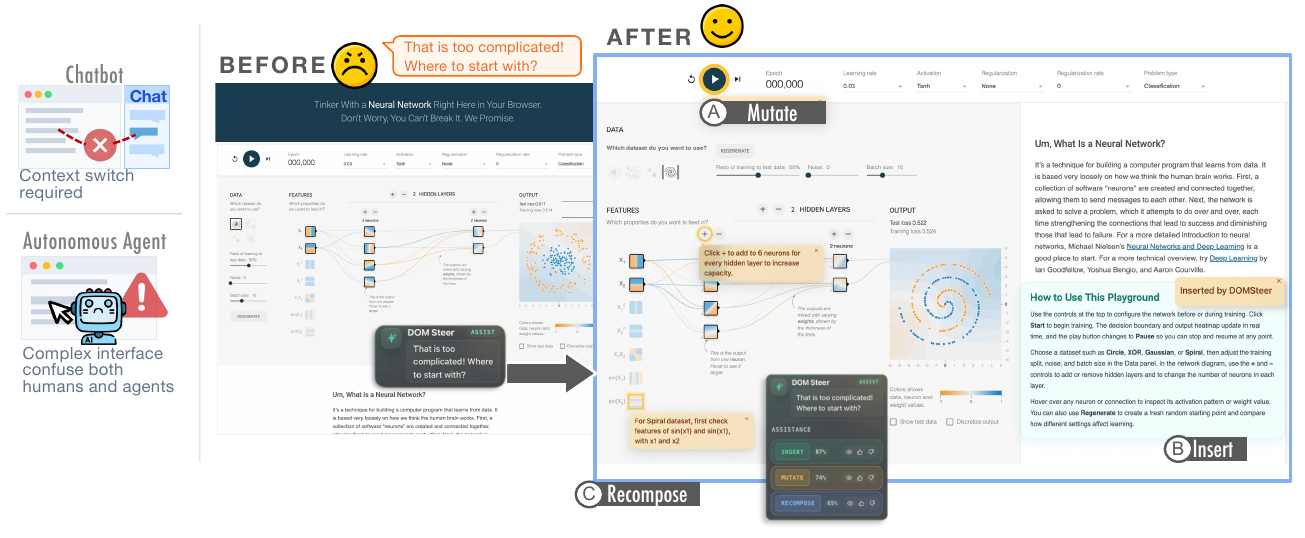}
  \caption{\sysName{} delivers assistance directly within the live interface through three DOM operations: (A) Mutate that adapts existing elements, (B) Insert that adds new contextual content, and (C) Recompose that reorganizes interface structure.}
  \Description{}
  \label{fig:teaser}
\end{teaserfigure}

\maketitle

\section{Introduction}
Graphical user interfaces (GUIs) have long been the dominant medium for understanding and interacting with data, ranging from chart creation~\cite{2017-voyager2}, visual analytics~\cite{ceneda2017vaguidance}, to communicative data dashboard~\cite{wang2018narvis}.
Many such tools have become de facto standards within their respective domains, deeply embedded in professional workflows, such as Autodesk in engineering design, Tableau in business analytics, and Epic in clinical healthcare.
However, the complexity of such interfaces is inherently coupled to the scale and complexity of the data they are designed to support.
As data grows in scale and complexity, these interfaces tend to expose users to dense, hierarchically organized visual components and interactions, resulting in steep learning curves that can overwhelm novices and reduce working efficiency.
 

Much effort has been devoted to mitigating these challenges, spanning interface tutorials and onboarding experiences~\cite{wang2018narvis, stoiber2022comparative}, and adaptive interfaces that adjust layouts, widget types, and toolbars based on user tasks, abilities, and usage patterns such as recency and frequency~\cite{Gajos2004SUPPLE-automaticUI, todi2018restructuring-layouts, Gajos2006designspace-adaptivegui, gobert2019self-adapting-menus, gajos2008ability-based-interfaces, wobbrock2011ability-based-design}. While these approaches offer valuable improvements, they are largely grounded in design heuristics or limited to predefined adaptation rules, constraining their ability to generalize to the diverse, context-dependent needs of real-world users at scale.

Recent advances in large language models (LLMs) have created new opportunities to address these challenges. 
One line of work augments GUIs with a separate conversational panel that provides on-demand guidance~\cite{weng2024insightlens, khurana2025guidedcopilot, chen2025interchat}. However, users often struggle to articulate the full interface context in language~\cite{jd2023johnny-cant-prompt}, and must further translate the system’s response back into the spatial and interactive structure of the GUI. Recent agentic systems, such as Gemini in Chrome~\cite{google_gemini_in_chrome} and ChatGPT Atlas~\cite{openai_chatgpt_atlas}, can respond to user requests through low-level interface actions such as clicking, typing, and scrolling~\cite{huq2025cowpilot}. 
However, their performance degrades on complex professional environments involving careful decision-making
(lower than 20\% accuracy for domain specific workflows~\cite{sun2025scienceboarde}).
%
Another line of work uses LLMs to dynamically generate task-specific interfaces tailored to user needs. For example, DynaVis~\cite{vaithilingam2024dynavis} generates interactive widgets for visualization editing based on natural language.
BISCUIT~\cite{cheng2024biscuits-ephemeralUI} creates ephemeral UI in computational notebook to understand and steer codes.
However, these generative interfaces often require constructing new interactive systems from scratch (\eg, rebuilding the connections to underlying data and reimplementing interaction logic).
This imposes substantial engineering overhead, underutilizes the mature interfaces already widely adopted in practice, and often assumes access to underlying data that may not always be available.

To address the gaps, we propose in-situ assistance that helps users navigate complex visual interfaces.
By \emph{in-situ assistance}, we refer to assistance delivered directly within the existing web-based interface that the user is actively working with, rather than through a separate interface or by rebuilding the application.
We enable this capability through browser-level interventions on the Document Object Model (DOM)~\cite{w3cdomlevel1}, allowing a client-side layer to add, modify, and interact with existing interface components without access or changing server-side logic.
%
To operationalize this idea, we propose a design space and a computational pipeline for in-situ assistance and then develop \sysName{}, a Chrome browser extension that delivers such assistance directly within live interfaces.
We evaluate \sysName{} on 101 annotated challenges across two real-world interfaces, DataVoyager and TensorFlow Playground. Our handbook-based retrieval achieves comparable assistance quality to direct LLM generation while reducing latency by 11.8$\times$. A within-subject user study ($N{=}12$) further shows that \sysName{} reduces task completion time by 25\% and improves task accuracy to 100\%, compared with 91.7\% for a chat-based baseline. 
%
Main contributions of this paper include
\footnote{\raggedright Project website and artifacts are available at
\url{https://visual-intelligence-umn.github.io/DOMSteer/}.}:

\begin{itemize}[leftmargin=*, nosep]
  \item A design space for in-situ assistance in web-based GUIs via direct DOM manipulation.

  \item A computational pipeline that efficiently and accurately translates user needs into live interface transformation.
  
  \item \sysName{}, a Chrome Browser extension that 
  operationalizes this pipeline to 
  deliver in-situ assistance for any web-based interfaces without needing access to their source code.

  \item A mixed-method evaluation of the proposed approach, , combining quantitative experiments with user studies to examine the effectiveness and usability of \sysName{}.
\end{itemize}

\section{Related Work}

\subsection{Adaptive and Dynamic User Interfaces}

Adaptive user interfaces have long been a long-pursuit goal in human–computer interaction, aiming to improve usability and task performance by tailoring interaction to users, tasks, and context~\cite{maes1993learning-interface-agents, horvitz1999mixed-initiative, Jameson200adaptive-interfaces-agents}. 
Prior work has explored various ways to design adaptive interfaces, 
such as adaptive layouts (\eg, grouping, ordering) and widget types to match users’ abilities or interaction histories~\cite{Gajos2004SUPPLE-automaticUI,todi2018restructuring-layouts,wobbrock2011ability-based-design}, adaptive menus and toolbars that promote, reorder, or highlight options based on usage patterns (\eg, recency, frequency)~\cite{Gajos2006designspace-adaptivegui,gobert2019self-adapting-menus} and dynamic control size and placement to reduce motor effort~\cite{gajos2008predictability}. 

More recently, LLM-powered systems have extended adaptive UI beyond rearranging existing controls, synthesizing task-specific, ephemeral elements from natural-language requests and interface context on demand~\cite{vaithilingam2024dynavis,cheng2024biscuits-ephemeralUI}, such as DynaVis~\cite{vaithilingam2024dynavis} adds interactive widgets for users to refine a visualization, BISCUIT~\cite{cheng2024biscuits-ephemeralUI} creates ephemeral UI scaffolds within computational notebooks to help users understand and steer codes.
Our work builds on the same goal of interface-embedded adaptation but shifts the delivery mechanism to the browser: we deliver assistance as reversible, transient DOM augmentation anchored to specific elements, enabling in-situ guidance that facilitate user navigation and can be deployed across existing heterogeneous web applications without modifying the underlying codebase.

\subsection{LLM-powered Interface Assistance}
Despite the growing adoption of large language models in interactive systems\cite{weng2024insightlens}, recent work highlights persistent cognitive barriers in prompt- and chat-based interaction \cite{khurana2024prompt-based-software-helpseeking, kiani2019software-newcomers}.
Prompting LLMs effectively is difficult \cite{jd2023johnny-cant-prompt, subramonyam2024gulf-envisioning}: users struggle to externalize intent that is spatial or dependent on interface state,
and underspecified prompts often lead to generic or overly verbose responses.
The resulting guidance then increases mental burden in translating and applying text-heavy outputs into actionable steps within the interfaces \cite{ khurana2024prompt-based-software-helpseeking, cranshaw2017calendar-help}. 
To reduce these barriers, recent systems increase relevance and actionability by leveraging context signals and tighter couplings to the interface. 
Some infer user intent from interaction traces, behavioral signals, or application state to provide context-aware assistance~\cite{zhao2025proactiveva,shaikh2025generalusermodel,lam2025justintime}. 
Others make assistance directly executable through click-to-apply edits on shared canvas~\cite{shin2025postermate, arjun2025pluto}, 
multimodal interaction that combine direct manipulation with natural language to enhance clarity~\cite{chen2025interchat, huang2025sketchGPT}, and step-by-step visual guidance for feature-rich software learning \cite{khurana2025guidedcopilot}.  

A common constraint is still that these assistants are engineered within specific interfaces: triggers, references, and action spaces rely on app-specific instrumentation and internal state models. 
As a result, assistance logic must be rebuilt for each new tool, and users who work across multiple web-based interfaces (\eg, data analysis, content authoring) cannot benefit from a unified experience. Our work follows the longstanding HCI goal of delivering situated, actionable, and context-aware assistance \cite{horvitz1999mixed-initiative, maes1993learning-interface-agents, amershi2019guidelines-human-ai}, but shifts the delivery mechanism to an lightweight, application-agnostic interaction layer at the web browser. 
We generate element-grounded DOM assistance over the shared DOM structure of web interfaces, enabling in-situ help without per-application rebuild.

\subsection{GUI/Web Agents}

GUI and web agents are designed to interpret interface elements (e.g., buttons, text fields, and menus) and act on them through direct interactions such as clicking, typing, and scrolling~\cite{nguyen2025guiagentssurvey}. Recent advances have shown promising results on routine, well-specified tasks such as form filling, online shopping, and search~\cite{zhou2024webarena, google_gemini_in_chrome, openai_chatgpt_atlas}. Realistic benchmarks, including WebArena~\cite{zhou2024webarena} and OSWorld~\cite{xie2024osworld}, have further accelerated progress in this area. Despite these gains, current agents remain far less effective on complex professional interfaces: the state-of-the-art model reaches only \textbf{78\%} accuracy on the WebArena Benchmark~\cite{webarena-leaderboard}, and the performance further drops to roughly \textbf{20\%} on interfaces that demand domain-specific judgment and multi-step reasoning over dense layouts~\cite{sun2025scienceboarde}. 
To cope with the imperfect performance,
some existing approaches allow agents to defer uncertain decisions to users through confirmation requests~\cite{peng2025morae-UIagent, huq2025cowpilot}. However, \textbf{deferring control to humans is insufficient when the interface itself is difficult for both agents and users to navigate}~\cite{cuvin2025decepticon}. 

To fill this gap, we propose in-situ assistance, which directly augments complex interfaces to make them more understandable and more supportive of task performance.

\section{Formative Study}
\label{sec:formative}
To characterize user challenges in complex visual interfaces, we conducted a formative study in which participants completed tasks on two representative interactive visual interfaces.
\subsection{Study Procedure}
We recruited 9 participants (5 female, 4 male; ages 21--30) who had general computer proficiency and no prior experience with the interfaces used in the study. 
The study protocol was reviewed and approved as exempt by the authors’ institution. Each participant received \$10 as compensation.
Participants completed exploratory and analytical tasks on two web-based 
visual interfaces:
\begin{itemize}[nosep,leftmargin=*]
\item 
\textbf{DataVoyager~2}~\cite{2017-voyager2}, a visual data exploration tool. Six tasks progressed from single-field lookups (\eg, identifying the highest-grossing genre) to multi-attribute analyses (\eg, whether commercial success relates to audience ratings).
\item 
\textbf{TensorFlow Playground}~\cite{smilkov2017visualization-deep-networks}, an interactive neural network visualization. Three tasks required participants to build, train, and evaluate neural networks across different classification and regression datasets.
\end{itemize}

Tasks were designed to cover the primary interaction demands of each interface. Full task descriptions are provided in the Appendix~\ref{formative:tasks}.
We used a think-aloud protocol and recorded participants’ on-screen interactions and their verbalized questions and comments. Each session lasted 30 minutes. 
The first author observed the sessions and provided only minimal clarification of task instructions without offering guidance on how to complete the tasks.
 
\subsection{Findings}
From the think-aloud transcripts and observation notes, we extracted 159 discrete instances in which a participant encountered a question or difficulty they could not resolve independently. 
The first author conducted an iterative qualitative analysis of these instances in a labeler interface (\autoref{fig:labeler}), which yielded six recurring challenge types: \textit{What}, \textit{Where}, \textit{How}, \textit{Why}, \textit{Next}, and \textit{Capability limits}. 
The emerging codes and category definitions were discussed with the research team and examined by the other authors throughout the analysis. 
Each category captures a distinct implicit question participants appeared to face at the moment of becoming stuck.

\paragraph{\textit{\textbf{WHAT} is this?} (n=40)} 
\label{challenge:what}
Participants struggled to interpret the meaning of interface elements, features, and terminology. The interface exposes elements that were unfamiliar with users, such as \textit{Wildcards} and \textit{Related Views} in Voyager 2; and domain-oriented concepts such as ``sigmoid activation'' and ``regularization rate'' in TensorFlow Playground. 


\paragraph{\textit{\textbf{WHERE} is it?} (n=31)} 
Participants had  specific element or concept in mind to take a certain action, but fail to locate it in the complex layout of an interface.
For example, 
P4 could not find and add the \textit{Production Budget} field to an encoding shelf in the long list; 
P1 knew the chart axis needed to be changed to log scale but dismissed the relevant control; P5 wanted to start training in TensorFlow Playground but could not locate the run button. 

\paragraph{\textit{\textbf{HOW} do I do this?} (n=25)} 
Even when participants had located the relevant UI elements, they may still struggle with the procedures required to complete the action, particularly when the interaction involved hidden, multi-step operations rather than a single click.
It involved either operational uncertainty about how to use the interface or interpretive uncertainty about how to interpret the outputs.
For example, P1 found the filtering area, but instead of dragging attribute names into the filter shelf to create a filter, they tried clicking the attribute names. Although P6 identified a scatter plot for anomaly analysis, they could not confidently determine which points qualified as outliers in this scatter plot.


\paragraph{\textit{\textbf{WHY} did it happen?} (n=23)} 
Participants may encounter unexpected interface behaviors that they could not readily explain.
For example, some participants were confused by why Voyager already showed many charts before they had explicitly constructed one, or why certain encoding options such as \textit{Column} are unavailable
sometimes.
In TensorFlow Playground, participants similarly questioned why a neural network configuration that worked on one dataset but failed on another, or why the loss curve was sometimes smooth and sometimes highly unstable. 

\paragraph{\textit{What should I do \textbf{NEXT}?} (n=25)} Participants also frequently reached moments where they did not know what actions to take to continue the task.
For instance, after constructing an initial chart in DataVoyager, some participants did not know whether to refine the current view or explore alternative charts to deepen the analysis.
In TensorFlow Playground, participants who observed that the model was not converging did not know whether to add layers, change the activation function, or try different input features.

\paragraph{\textit{\textbf{CAN} the tool support this?} (n=15)} Finally, some participants wanted to perform operations that were not fully supported by the interface.
For example, several participants wanted to compute derived values in DataVoyager, such as return on investment defined as gross divided by budget, but the interface does not support calculated fields.
In TensorFlow Playground, participants wanted a mechanism to track and compare previous model configurations during trial-and-error exploration, which are not supported.


\noindent
{\bf Summary:}
Across both interfaces, the instances of friction span from surface-level recognition failures to higher-level exploration dead ends. 
These challenges were tightly coupled with the particular elements and contingent on a particular interface state.
Such situatedness suggests that effective assistance for complex interfaces may need to be similarly grounded, delivered at the relevant element and responsive to the current state. 
In the next section, we build on this observation to explore a design space for in-situ assistance delivered through DOM-based interface manipulations.
\section{A Design Space for In-situ Assistance}


\begin{figure}[t]
 \centering
\includegraphics[width=\columnwidth]{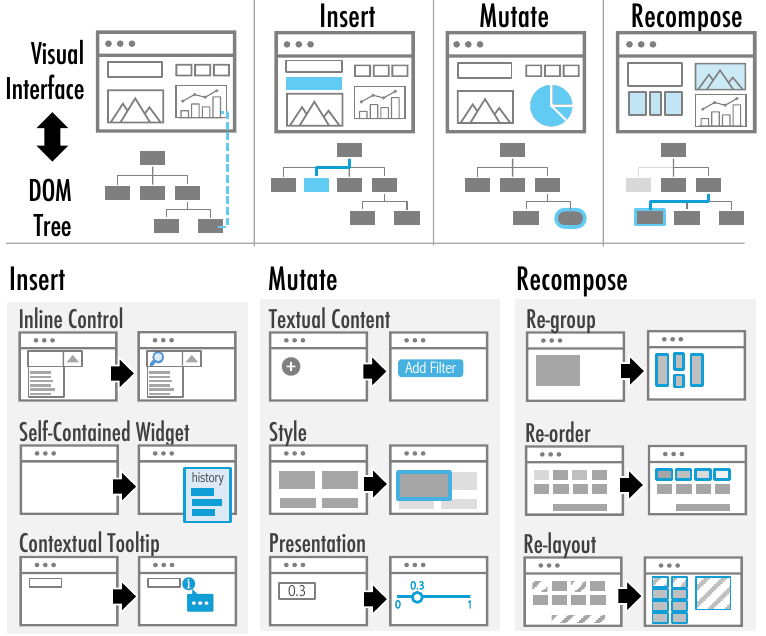}
  \caption{Design space of DOM-mediated in-situ assistance}
  \label{fig:dom-design-space}
\end{figure}

Our design space organizes in-situ assistance by how it intervenes in an existing interface to address different kinds of user challenges identified in the formative study. The three dimensions differ in the role they play for users. \textbf{Insert} adds new assistance content when the original interface does not provide enough local guidance. \textbf{Mutate} modifies existing interface elements to make them easier to notice, understand, or operate. \textbf{Recompose} reorganizes existing elements to better match the user’s task and reduce navigational burden.
We intentionally exclude \textbf{removal} as a intervention type to preserve the original interface’s full functionality. Elements that are irrelevant in the current context can instead be deemphasized through mutation.

Note that the design dimensions are not mutually exclusive. The same assistance goal may be achieved through different intervention types depending on how the interface is modified. For example, making a control easier to notice could be achieved by \textit{mutating} its visual style (\eg, border) to increase salience, or by \textit{inserting} a tooltip or annotation that calls attention to it.


\subsection{Insert}
\textbf{Insert} operations address user challenges that arise because the sufficient support are not provided at the point of need. 
Adding additional support has been a common strategies to improve the usability and effectiveness of an existing interface~\cite{lafreniere2013enhanced-tutorials, metajka2011-ambient-help, grossman2010-toolClips-video-edit-assistance}.
While insertion assistance can be useful for all categories of user struggles, it is especially helpful when participants needed help interpreting an unfamiliar UI components (WHAT), understanding an unexpected system behavior (WHY), determining how to perform a multi-step action (How), or accessing new interactive support the interface does not natively provide (CAN).

Insert assistance is implemented by creating new DOM nodes at runtime and attaching them to the live page. A generated script calls \texttt{document.createElement()} to construct the new element, then places it using one of two strategies depending on the desired layout behavior. When the new element should flow naturally within the surrounding structure (\ie, inheriting the existing spacing and alignment), it is inserted adjacent to the target via \texttt{appendChild()}, \texttt{insertBefore()}, or \texttt{insertAdjacentElement()}. When tight coupling to the target's local layout is undesirable, the element is instead appended higher in the tree (typically \texttt{document.body}) and visually anchored to the target through absolute positioning.

We identify three representative subtypes of insert.
\textbf{Contextual Tooltips} add ephemeral overlays anchored to a target element to provide local explanation of UI components (WHAT) or unexpected system behaviors (WHY). 
For example, in Voyager2, a tooltip near the quantitative wildcard can explain WHAT the wildcard represents or WHY it is not responding when users click. 
\textbf{Inline Controls} address CAN and HOW challenges by inserting new interactive elements directly into an existing interface region. Inline controls are tightly coupled to existing interface components by extending or streamlining interactions that are already present, 
such as a search input inserted at the top of a long dropdown list, or a sorting toggle added within a table header that collapses a multi-step menu interaction into a single click.
\textbf{Self-contained Widgets} address CAN challenges by adding independent, floating components that collect and present information without interacting with existing interface elements. For example, a floating panel that records and compares alternative configurations during trial-and-error exploration, or a side panel that summarizes the current chart state and suggests possible next analytical steps.

\subsection{Mutate}
\textbf{Mutate} operations address user challenges that arise not from missing functionality, but from the difficulties to effectively locate (WHERE), understand (WHAT), or operate these existing functionalities (HOW). 
Adapting existing interface elements to better match user context and expectations is a well-established strategy for improving usability~\cite{Gajos2004SUPPLE-automaticUI, agarwal2013-widgetlens, Baudisch2006Phosphor-afterflow-effects}. 


Mutate operations are implemented by directly modifying the properties of existing DOM nodes, including their text content, visual styling, or interaction attributes, without adding new elements or restructuring the page. 
Concretely, this covers operations such as replacing an element's content via \texttt{element.textContent}, altering its appearance through \texttt{element.style} or \texttt{element.classList}, or changing an input attribute to expose a different control modality. This allows the assistance to reshape how an element reads, looks, or behaves while leaving the surrounding interface intact.

We identify three representative subtypes of mutate.
\textbf{Text Content} mutations address WHAT challenges by rewriting labels, descriptions, or placeholders to improve comprehension.
For example, a small button labeled ``+'' may have weak affordance, and rewriting it as ``Add Filter'' can make the function more interpretable.
\textbf{Style} mutations address WHERE challenges by modifying the visual presentation of an existing element, such as color, opacity, border, size, or animation, to direct attention toward what is relevant in the user's current context. 
For example, one assistance may highlight the sort control among several nearby icons, or visually emphasize a chart region that contains potential outliers.
%
\textbf{Presentation} mutations address HOW challenges by changing how an existing function is exposed or manipulated while preserving its underlying role in the interface. 
For example, an RGB-number color field may be converted into a color picker, and a numeric threshold input may be replaced with a slider embedded directly in the visualization.

\subsection{Recompose}
\textbf{Recompose} operations address user challenges that arise not from missing or poorly presented individual elements, but from a spatial or structural arrangement 
that make relevant controls scattered across the page (WHERE), related items lack visual grouping (WHAT), or step-by-step work unnecessarily difficult (NEXT).
Reorganizing the structural and spatial relationships among existing interface elements is a well-established strategy for reducing interaction friction and improving task alignment~\cite{Gajos2004SUPPLE-automaticUI, Gajos2006designspace-adaptivegui, findlater2004static-adaptive-adaptable-menus}.

At the implementation level, recompose first identifies the target interface elements and then reorganizes their structural relationships in the live DOM. 
Target identification can rely on standard DOM selection APIs (\eg, \texttt{querySelector()}, \texttt{getElementById()}) and traversal methods (\eg, \texttt{parentElement}, \texttt{children}, or \texttt{closest}).
Once identified, existing nodes can be rearranged by moving them into the target parent container with APIs such as \texttt{appendChild()} and \texttt{insertBefore()}. 
We identify two representative subtypes of recompose.
\textbf{Re-Group} addresses WHAT and WHERE challenges by introducing or strengthening parental structure so that related elements are visually and spatially clustered, making semantic relationships more explicit. For example, data fields in a movie dataset may be grouped into named categories such as ``Audience Feedback,'' ``Production,'' and ``Commercial Results,'' reducing visual search and helping users understand how controls relate to one another.
\textbf{Re-Order} addresses WHERE and NEXT challenges at a local level by changing the sequence of elements within a container. For example, a long list of analysis options may be reordered to surface the most relevant attributes first.
\textbf{Re-Layout} addresses WHERE and NEXT challenges at a global level by repositioning entire interface regions relative to one another. 
For example, different views can be reordered to reflect the temporal sequence in which they are typically needed for finishing a task, reducing unnecessary navigation.

\section{In-Situ Assistance from \sysName{}}

\subsection{System Overview}
\begin{figure}[t]
 \centering
\includegraphics[width=\linewidth]{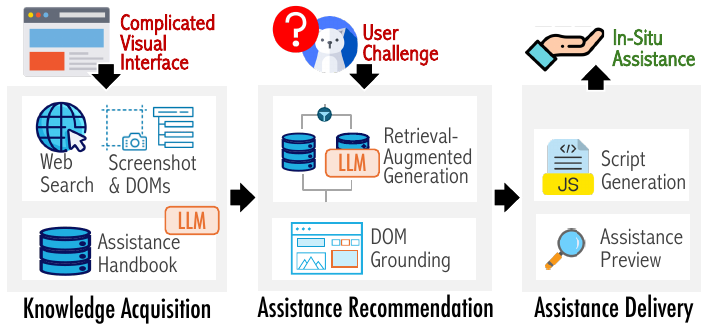}
  \caption{Computational pipeline of \sysName{}. 
  }
  \label{fig:pipeline}
\end{figure}

As shown in \autoref{fig:pipeline}, \sysName{} consists of three key modules: Knowledge Acquisition, Assistance Recommendation, and Assistance Delivery. 
For any web-based interface, users can invoke \sysName{}, implemented as a lightweight 
Chrome Manifest V3 extension that appears as a floating overlay on the active page. 
Once activated, \sysName{} initializes itself via the Knowledge Acquisition module, constructing a lightweight knowledge 
base about the current interface using both screenshots and web search on official tutorials and documentations.
Users can continue interacting with the interface as usual.
At any point, users can open the floating overlay and describe the challenge they face in natural language, optionally referencing specific interface elements to guide the response.
Using the interaction history and interface knowledge base, the Assistance Recommendation module recommends the top three assistance candidates, each grounded in specific DOM elements in the current interface. 
Based on the grounded elements and the selected suggestion, the Assistance Delivery module
directly injects DOM modifications into the interface, while also allowing users to 
preview and select alternative high-ranking options. 
Users can also provide feedback on the assistance, which incrementally improves recommendation quality.
In the remainder of this section, we describe each module in detail.

\subsection{Knowledge Acquisition}
When \sysName{} is first activated on a new interface, it constructs a lightweight 
knowledge base comprising three components: a summary of external documentation, a 
structured representation of the live interface, and a pre-computed handbook of assistance 
cases for efficient, reliable assistance generation at runtime.

For external sources, \sysName{} queries Tavily web search API~\cite{tavily-search} to retrieve online tutorials, official documentation, and feature descriptions for the target interface. 
The retrieved documents are summarized by an LLM (selectable from 
a provided list of LLM models) into a structured markdown file, capturing: what the interface is, what features it exposes, and what interactions it does or does not supports.
Although knowledge base construction takes 3–5 minutes on average, it is a one-time cost and the resulting knowledge base is reused across all subsequent sessions on the same interface.

For the live interface, \sysName{} stores the screenshot of the current interface, and programmatically identifies all interactable elements on the page (\eg, buttons, dropdowns, sliders, and interactive visualizations) directly from the page's DOM tree, requiring no computer vision or LLM involvement.
The DOM identification is based on the implementation in WebArena~\cite{zhou2024webarena}.
Each DOM is assigned a unique index and a corresponding bounding box to make them interpretable and referenceable in downstream reasoning.

While it is possible for LLMs to directly use the above two sources, combined with the provided user challenges, to generate assistance, we observed long latency in our preliminary measurements. Specifically, generating assistance via a frontier LLM (\eg, GPT-4o) takes between 4--20 seconds per request depending on interface complexity, far exceeding the response latency acceptable for interactive use.
%
Therefore, we construct a set of $N$ representative assistance cases as an 
\textit{Assistance Handbook} for reliable and efficient assistance generation in ad. 
We set $N=120$ based on empirical experiments on DataVoyager and TensorFlow Playground, balancing pre-processing overhead against on-the-fly generation latency.
In our implementation, generating these 120 assistance cases takes about 95 seconds during initialization.
Each case stores five fields: 
(1) the \textit{Assistance} to provide, expressed in natural language; 
(2) the \textit{DOM in-situ assistance subtype}, defined according to our design space (\autoref{fig:dom-design-space});
(3) the \textit{UI Targets} list, where each target is described by its semantic meaning rather than brittle DOM indices that may vary across users, interface states, screen sizes, and viewing conditions;
(4) the \textit{Rationale}, which explains the underlying challenge that the assistance is intended to mitigate; and 
(5) the \textit{Configuration}, which specifies the parameters needed to instantiate the intervention on the page.
The handbook is generated by prompting a user-selected LLM with the interface knowledge (from both the web search and the live interface) and the full design space (prompts and examples in \ref{prompt:generation}). 
The resulting cases are embedded using \texttt{text-embedding-3-small} and stored in an 
in-memory FAISS index, enabling efficient cosine-similarity retrieval.


\subsection{Assistance Recommendation}

Given a challenge described by users, \sysName{} follows a two-step process to generate assistance. 
It first identifies the most relevant assistance from the handbook or generates a new one on the fly, then grounds the recommended assistance to DOM elements.

\textbf{Retrieval-Augmented Generation.}
For a given user challenge, \sysName{} generates its embedding using 
\texttt{text-embedding-3-small} and searches the Assistance Handbook for the top-$k$ 
($k{=}3$) most relevant cases via cosine similarity against the \textit{Rationale} 
field of each stored case.
If the highest-scoring case exceeds a retrieval threshold $\tau=0.5$ (empirically set based on pilot testing), the system reuses that Handbook case directly, 
serving the response in low latency without invoking the LLM.
Otherwise, it falls back to synthesizing a new assistance case on the fly using the 
same \texttt{gpt-4o}-based generation procedure as the handbook construction stage.
The newly generated case is added to the handbook for future reuse.

Note that for both retrieved or freshly generated assistance, they stores UI Targets as 
abstract semantic identifiers rather than concrete DOM references, since DOM paths are brittle and vary across users, screen resolutions, and interface states.
Therefore, before the recommended assistance can be delivered, its semantic target descriptions must be linked to the correct UI elements in the current interface, a step we call DOM grounding.

\textbf{DOM Grounding}.
DOM grounding leverages DOM identification of the current interface (Knowledge Acquisition), semantic descriptors of UI targets in the assistance (Retrieval-Augmented Generation) that are already computed, incurring no additional LLM calls or DOM traversal at this stage.
Specifically, for each UI target in the recommended assistance, it semantic description is 
matched to the DOM element whose embedding has the highest cosine similarity, 
resolving all targets to concrete element indices ready for delivery.

\begin{figure}
    \centering
    \includegraphics[width=\linewidth]{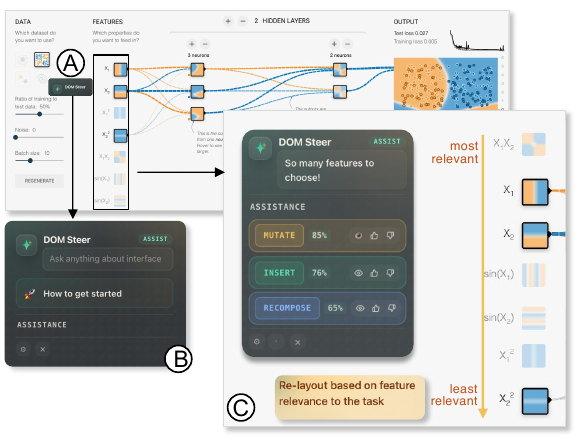}
    \caption{ \sysName{} applied to TensorFlow Playground.}
    \label{fig:interface}
\end{figure}

\subsection{Assistance Delivery}

Once all targets of recommended assistance are grounded to concrete DOM element indices, \sysName{} materializes the assistance directly in the live interface.

Each assistance \textit{type} in the design space (\autoref{fig:dom-design-space}) 
maps to a pre-implemented JavaScript delivery script selected by the assistance \textit{subtype} and parameterized by the assistance 
\textit{configuration}.
For example, inserting \textit{Contextual Tooltip} is realized by injecting a floating \texttt{<div>} anchored to the target element's bounding box via absolute CSS positioning, whereas mutating \textit{Style} is applied by directly overriding the target element's CSS class list.
\textit{Widget} inserts a self-contained HTML container that renders the React components specified in the configuration of the assistance.
All DOM operations are executed by the content script via Chrome extensions API, which injects and runs the delivery script in the context of the current interface.

In \sysName{}, users can easily pre-view, remove, or accept recommended assistance, as shown in \autoref{fig:interface}.
To support reversibility, every element created or modified during Assistance Delivery is tagged with a special attribute at injection time.
This allows \sysName{} to cleanly identify, update, or remove any previously delivered 
assistance at any point without altering the underlying application state.

\subsection{Implementation}
\sysName{} is built on top of Nanobrowser~\cite{nanobrowser2025}, an open-source 
Chrome Manifest V3 extension for multi-agent web automation, which provides the 
foundational infrastructure for browser interaction, DOM access, and multi-agent 
coordination.
We extend its architecture with our handbook-based 
assistance pipeline and DOM modification delivery scripts.
The extension is implemented in TypeScript with a React~18 and Tailwind CSS, rendered in a shadow DOM to isolate extension styles from the host page.
LLM integration is handled through LangGraph~\cite{langgraph2024}, supporting multiple providers (OpenAI, Anthropic, Gemini, Ollama) so users can configure different models for 
different pipeline stages.

\section{Assistance Quality and Efficiency Evaluation}
In this section, we assess the quality and efficiency of \sysName{} in the generation of in-situ assistance.

\subsubsection*{\bf Evaluation Dataset.}
We constructed a benchmark dataset using real user data collected in the formative study (\autoref{sec:formative}). 
Specifically, the dataset comprises 101 annotated user challenges spanning two complex web interfaces (DataVoyager and TensorFlow Playground) with coverage across all six challenge categories defined in our design space.
Each entry consists of three components: 
(1) the user's challenge description in natural language, 
(2) the interface state (DOM tree) captured at the moment the challenge was reported, 
and (3) manually annotated example assistances that resolve the challenge. 
Annotations were produced by one author and reviewed by two additional co-authors. Importantly, the annotated assistances are not treated as exhaustive ground truth, 
since a single challenge may admit multiple valid resolutions, but rather as validated reference examples for evaluation purposes.



\subsubsection*{\bf Evaluation Methods and Metrics.}
We evaluated \sysName{} against two other methods.
\textit{Baseline: LLM Generation} generates assistance on the fly from an LLM based on the user need, the current interface DOM, and the interface knowledge base. 
\textit{Our Method 1: Handbook Retrieval} retrieves the most similar assistance case from the pre-generated assistance handbook. 
\textit{Our Method 2: Handbook Retrieval with Fallback} (Handbook + FB) uses a hybrid strategy that first retrieves from the handbook and falls back to LLM Generation when the top retrieval score falls below a confidence threshold of 0.5 selected empirically from pilot inspection as a conservative cutoff for low-confidence retrievals.

\begin{figure}
    \centering
    \includegraphics[width=\linewidth]{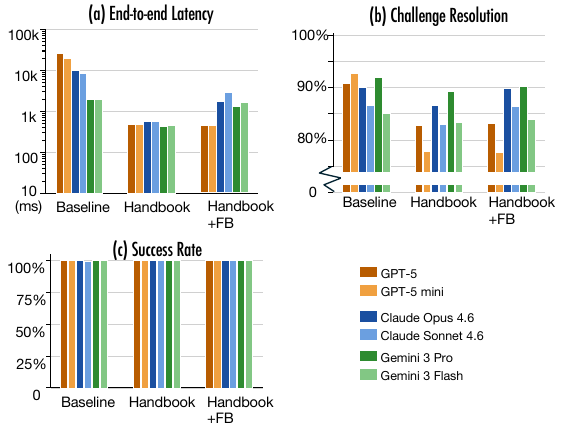}
    \vspace{-2em}
    \caption{Evaluation of LLMs across three methods. Zigzag markers indicate truncated y-axes. Darker shades denote larger LLM models.}
    \label{fig:performance}
\end{figure}

Each method returns a browser-executable script that implements in-situ assistance. 
We evaluate the generated assistance using three metrics: 
(1) \textit{Success Rate}, the proportion of cases in which the returned script executes without errors; 
(2) \textit{Latency}, the end-to-end response time; and (3) \textit{Challenge Resolution}, the degree to which the rendered assistance addresses the user's challenge. For Challenge Resolution, we employ LLM-as-Judge evaluation using GPT-5.4, following the now-standard practice in open-ended web agents where human annotation does not scale~\cite{webarena-leaderboard, zhou2024webarena}. The manually annotated example assistance is provided to the judge as a reference, grounding LLM-as-Judge assessment in human-validated expectations (prompt in~\ref{prompt:judger}). To examine robustness across backbone LLM choices, we evaluate each method using six large language model backbones spanning both larger and smaller variants from three families (GPT, Claude, and Gemini).


\subsubsection*{\bf Results Analysis.}
The results are summarized in \autoref{fig:performance}.
Baseline incurs substantially higher response times (a), on average \textbf{11.8 $\times$} slower than handbook-based methods. 
Notably, smaller models (GPT-5 mini, Claude Sonnet 4.6, Gemini 3 Flash), despite being designed for speed and cost efficiency, did not significantly reduce latency compared to their larger counterparts. We attribute this to the complexity of the task: less capable models require longer reasoning chains, suggesting that latency in the direct LLM generation cannot be mitigated simply by reducing model size.

These latency gains come with little sacrifice in other performance metrics.
For challenge resolution, Baseline performs best (mean 8.96/10), followed by Handbook+FB (8.52/10) and Handbook (8.38/10).
But the differences among three methods on challenge resolution are small, indicating strong task-solving capability regardless of pipeline configuration.
Across all methods and LLM backbones, success rates are uniformly high (mostly 100\% except one 99\% on baseline), confirming that LLMs are reliably capable of producing valid DOM modifications. 

Handbook and Handbook+FB achieve comparable performance across all metrics, reflecting strong handbook coverage. 
The fallback path is rarely triggered, activating in only 2\% of test cases. Importantly, the test set was held out entirely from the handbook generation process, confirming that the handbook generalizes to unseen challenges and captures a diverse and representative range of assistance scenarios.

\begin{figure*}
 \centering
\includegraphics[width=0.95\linewidth]{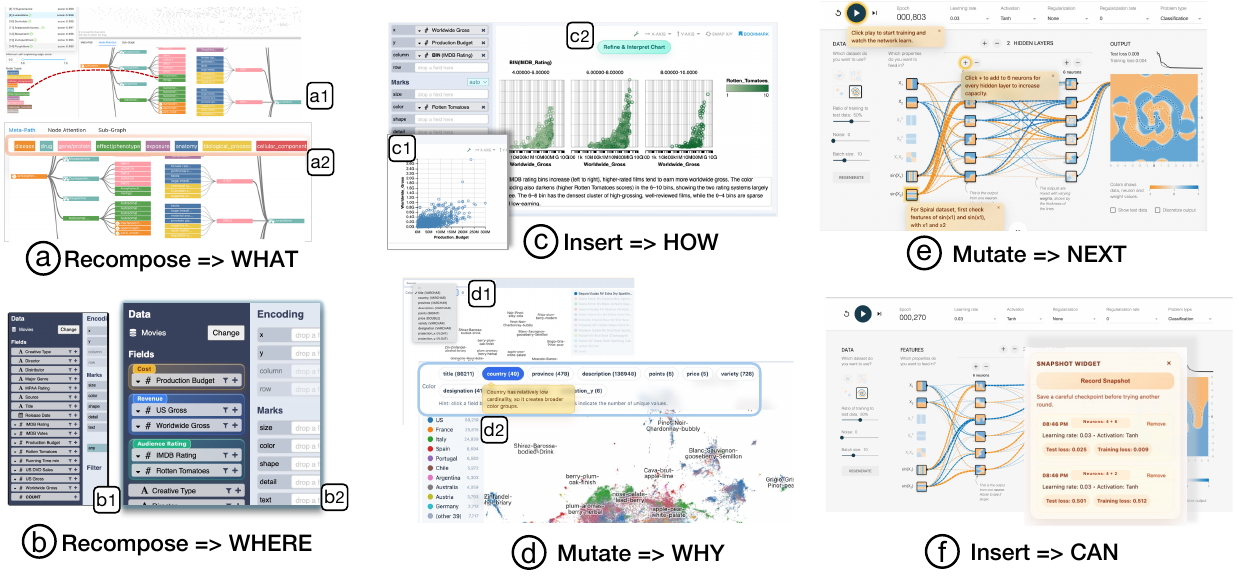}
  \caption{Representative use cases of \sysName{} across four web interfaces, organized by challenge type.}
  \label{fig:use-cases}
\end{figure*}

\section{Use Cases}
To illustrate how \sysName{} generalizes beyond the two interfaces used in evaluation, we present representative use cases organized by the six challenge types from our formative study. The use cases span four complex visual interfaces: DataVoyager~\cite{2017-voyager2} for tabular data exploration and visualization, TensorFlow Playground~\cite{smilkov2017visualization-deep-networks} for interactive machine learning, TxGNN Explorer~\cite{huang2024txgnn} for biomedical knowledge graph exploration, and Embedding Atlas \cite{ren2025embeddingatlas} for high-dimensional data visualization.

\textbf{WHAT.}
In TxGNN Explorer, users inspect relationships among medical entities (\eg, drug, disease, and gene) through multiple coordinated views. In one view, a user sees colored nodes and asks what the colors represent. Although a node-type legend is available, it is easy to overlook because it appears far from the graph at the bottom of the sidebar and in a separate interface region, weakening its perceived connection to the graph being interpreted (Figure~\ref{fig:use-cases}.a1). 
\sysName{} then \textbf{Recomposes} the interface by moving the legend onto the graph canvas for easier reference and reordering it into a more semantically meaningful sequence (Figure~\ref{fig:use-cases}.a2).


\textbf{WHERE.}
In DataVoyager, a user imports a movie dataset and wants to examine the relationship between audience feedback and commercial success.
However, DataVoyager by default shows all data fields from the imported datasets, making it hard for the user to locate relevant fields from a long list  (Figure~\ref{fig:use-cases}.b1). 
\sysName{} can \textbf{recompose} the interface by \textit{re-grouping} the relevant fields with semantic labels: \textit{Revenue} (Worldwide Gross, US Gross, US DVD Sales), \textit{Cost} (Production Budget), and \textit{Audience \& Critics} (IMDB Rating, Rotten Tomatoes Rating), so that related fields can be located quickly by concepts (Figure~\ref{fig:use-cases}.b2). 

\textbf{HOW.}
In DataVoyager, the user continues working on the movie dataset and builds a scatter plot of budget versus gross.
However, the user is unsure how to refine the chart to interpret the dense, skewed view. 
\sysName{} can \textbf{Insert} an inline ``Refine \& Interpret Chart'' control near the visualization (Fig.~\ref{fig:use-cases}.c), exposing shortcuts to switch to log scale and place a third field on \textit{Row} to split the chart into groups, while also generating a brief textual interpretation of the visible patterns for the chart. 

\textbf{WHY.}
In Embedding Atlas, a user imports a wine review dataset in which each review is encoded as a high-dimensional embedding and projected onto a scatter plot via dimensionality reduction. 
The user colors the plot by the \textit{description} field, after which most points disappear and only a handful visible. 
This because that \textit{description} contains thousands of unique values, far exceeding the number of colors the default encoding scheme can represent, making most points receive no color assignment. 
\sysName{} can \textbf{mutate} the field selector into a set of clearly labeled buttons, augmenting each label with a cardinality indicator (Figure~\ref{fig:use-cases}.d). 
This makes the color encoding process more intuitive and naturally steers users toward lower-cardinality fields that are better suited for color encoding.

\textbf{NEXT.}
After constructing an initial neural network in TensorFlow Playground, users may still be uncertain about how to proceed NEXT, particularly when many controls remain unused. 
In such cases, \sysName{} can \textbf{mutate} the \textit{styles} of these untouched controls (\eg, additional feature toggles and buttons for adding neurons) to make them more visually salient and guide users’ attention toward potentially relevant actions (Figure~\ref{fig:use-cases}.e). 
It can further \textbf{insert} tooltips that provide brief suggestions for possible next steps.

\textbf{CAN}
Also in TensorFlow Playground, a user may go through several rounds of adjusting features, hidden layers, and learning rates, yet the interface offers no history mechanism for preserving promising settings. 
In such cases,
\sysName{} can \textbf{insert} a ``Configuration Snapshots'' \textit{widget} near the output region with a save button, allowing the user to preserve the current configuration together with its results and revisit earlier alternatives (Figure~\ref{fig:use-cases}.f).

\section{Usability and Effectiveness Evaluation}
\label{sec:user-study}

This section evaluates the usability and effectiveness of \sysName{}.
Specifically, we first conducted a within-subject user study to evaluate \sysName{} against a chat-based web assistant on two different user interfaces. We then compared both systems with an autonomous web assistant on the same tasks.

\subsection{Study Setup}

\noindent
\textbf{Participants.}
We recruited 12 participants (8 female, 4 male; 6 aged 18--24 and 6 aged 25--34) with general computer proficiency and little prior experience with the interfaces used in the study. None had participated in the formative study. 
The study protocol was approved as exempt by the authors' institution. Each participant received \$10 in compensation.

\noindent
\textbf{Assistance Conditions.}
We evaluated \sysName{} against two baselines from ChatGPT Atlas~\cite{openai_chatgpt_atlas}, a state-of-the-art AI-native browser developed by OpenAI and among the most capable conversational web assistants publicly available.  
The \textit{chat-based assistant} provides help through a side-panel conversational interface with live access to the webpage, while leaving the interface itself unchanged. This served as the primary \textit{participant-facing} baseline.
The \textit{Atlas Agent Mode} is an autonomous agent that attempts to complete tasks directly without user intervention. We used this Agent Mode as \textit{a system-level baseline} to situate \sysName{} relative to end-to-end automation.


\noindent
\textbf{Tasks.}
We designed eight tasks across two web-based visual interfaces, ranging from simple lookup and navigation to more complex interpretation and design, and covering the challenge types identified in the formative study (\autoref{sec:formative}).
\begin{itemize}[leftmargin=*]
    \item \textit{DataVoyager~2.} Participants used the interface to explore a cars dataset~\cite{vega_datasets} and completed four tasks with ground-truth answers:
    (T1)~identifying the most fuel-efficient car,
    (T2)~determining which origin showed the widest horsepower range,
    (T3)~counting European cars with both horsepower above 100 and four cylinders, and
    (T4)~describing characteristics of eight-cylinder cars relative to other cylinder counts.
    \item \textit{TensorFlow Playground.} Participants completed four exploratory tasks:
    (T5)~locating and explaining the effect of the \textit{Discretize} toggle on the output visualization,
    (T6)~locating one misclassified test data point,
    (T7)~designing a neural network that performed well on two regression datasets, and
    (T8)~designing a neural network that performed well across four classification datasets.
\end{itemize}

\textbf{Procedure.}
The user study compared DOMSteer and the chat-based assistant. 
Each participant used both assistants, each paired with a different interface. We counterbalanced assistant--interface pairings and presentation order across four conditions using a balanced Latin square, with three participants per condition.
Participants first received a brief tutorial on the basic interaction mechanics of both assistants.
Participants were then instructed to solve the tasks as accurately as possible, think aloud throughout the session, and ask for clarification about task wording when needed.
After completion of all tasks, participants filled out a post-study usability questionnaire and took part in a short semi-structured interview. Session lasted approximately 30--40 minutes.

We collected both quantitative and qualitative data. 
For quantitative analysis, we measured task accuracy, completion time, and self-reported usability ratings. Task accuracy was scored against predefined task-specific ground-truth answers and criteria.
For qualitative analysis, we analyzed think-aloud transcripts, interaction observation, and post-study interview responses to understand how participants perceived the two assistants and how the delivery format of assistance shaped their experience.

\noindent
\textbf{Autonomous baseline procedure.}
We separately evaluated ChatGPT Atlas Agent Mode on the same eight tasks. The agent completed T1--T4 sequentially in a single DataVoyager session and T5--T8 sequentially in a single TensorFlow Playground session, allowing interface familiarity to accumulate across tasks. 
We repeated this procedure for six independent runs (48 task instances in total) and evaluated task correctness and completion time using the same rubrics as in the user study.

\begin{figure}
    \centering
    \includegraphics[width=1\linewidth]{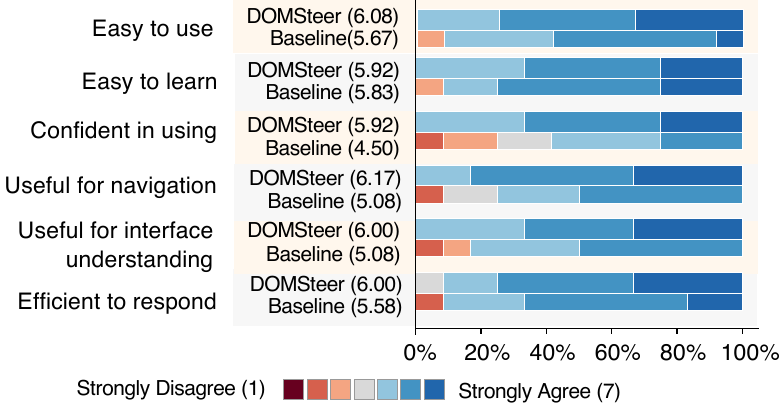}
    \caption{User Ratings. Comparison of \sysName and the baseline based on the post-study questionnaire responses. Participants rated each item on a 7-point Likert scaling ranging from strongly disagree (1) to strongly agree (7).}
    \label{fig:user-rating}
\end{figure}

\begin{figure}
    \centering
    \includegraphics[width=1\linewidth]{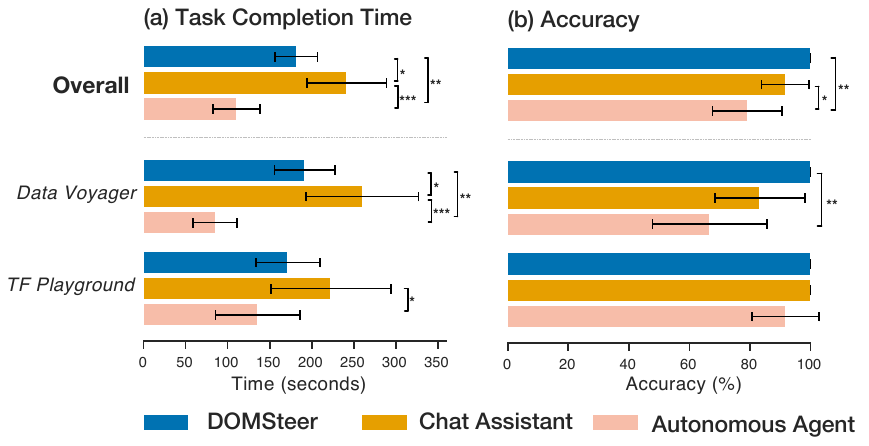}
    \caption{Task completion time (a) and accuracy (b) for \sysName{}, the Chat Assistant, and the Autonomous Agent, shown overall and by interface. Error bars indicate 95\% confidence intervals.
    }
    \label{fig:combined-time-acc}
\end{figure}

\subsection{Quantitative Results.}
We report three quantitative outcomes: post-study usability ratings, task completion time, and task accuracy. 
Usability ratings were collected only for the two participant-facing conditions (\sysName{} and the Chat Assistant), whereas completion time and task accuracy were compared across all three conditions.
We used Student’s t-tests to compare conditions on accuracy and time.


\noindent
\textbf{User Ratings.} 
\sysName{} led to improved user experience in navigating the complex visual interfaces.
Figure~\ref{fig:user-rating} presents participants' subjective ratings comparing \sysName{} and baseline across six dimensions related to the usability and utility. 
Overall, participants rated \sysName{} more positively than the baseline across all six questionnaire items. 
They also reported stronger agreement on its \textit{useful for navigation guidance} and \textit{useful for interface understanding}. 

\noindent
\textbf{Task Completion Time.}
As shown in \autoref{fig:combined-time-acc}a,
\sysName{} significantly reduced task completion time compared with the Chat Assistant ($t=2.44$, $p=.0161$).
It is not surprising that Autonomous Agent is significantly faster than \sysName{} ($t=-2.89$, $p=.004$) and Chat Assistant  ($t=-5.33$, $p=.0001$). 
Across all tasks, the overall mean completion time was 181.7$\pm$25.3 seconds for \sysName{} and 241.5$\pm$47.2 seconds for the baseline (95\% CI). This pattern held across both interfaces. In DataVoyager, mean completion times were 191.7$\pm$35.9 seconds for \sysName{} and 260.1$\pm$66.7 seconds for the chat assistant baseline (95\% CI). In TensorFlow Playground, the corresponding means were 171.6$\pm$37.9 seconds and 223.0$\pm$71.1 seconds (95\% CI). These results indicate that \sysName{} was faster than chat assistant across both interface contexts.

\noindent
\textbf{Task Accuracy.} 
As shown in \autoref{fig:combined-time-acc}b,
participants' performance in \sysName{} was significantly more accurate than automation ($t=3.56$, $p=.0005$), but no significance compared with baseline ($t=1.42$, $p=.156$).
Overall task accuracy was 100\% for \sysName{}, 91.7\% for the baseline, and 79.2\% for automation. 
These accuracy differences were driven primarily by DataVoyager, where \sysName{} significantly outperformed both the baseline and automation, while no significant differences were observed in TensorFlow Playground. 

\subsection{Qualitative Findings}
We analyzed think-aloud transcripts and post-study interviews to understand where the two systems differed in use. 

\paragraph{Chat-based assistance imposed translation and location burden.}
Because the baseline delivered guidance in a side panel separate from the interface, participants had to mentally bridge between textual instructions and the actual controls they referenced. 
In T1, for example, the chat assistant usually suggested sorting by miles per gallon to identify the most fuel-efficient car, but participants struggled to find the small, unlabeled sort icon while reading instructions in a different part of the screen. Some had created a bar chart listing all car names, but without successfully sorting the chart by MPG.
As the chart was long and only part of it was visible at once, they compared only the visible subset and selected the highest-MPG car within that partial view rather than in the full dataset. 
Participants who did not realize they could scroll through the chart therefore produced incorrect answers such as ``Audi 5000S.'' The problem was not that the guidance was unreasonable, but that its out-of-band delivery left the work of locating, executing, and verifying the procedure with the user. 
This translation burden explains why \sysName{}'s accuracy advantage was concentrated in DataVoyager: tasks requiring precise element-level interaction benefited more from in-situ guidance that eliminated the gap between instruction and execution.

\paragraph{Fluent reasoning sometimes replaced grounded verification.}
When participants directly prompted to get the answer to a task, the system could return a candidate answer together with a long explanation of how it had reasoned or how the result could be done for verification. 
In practice, two participants in T3 found it difficult to follow these verification steps back in the interface. As a result, some accepted the suggested answer on the basis of the detailed reasoning itself rather than on grounded confirmation, while the answer given was wrong. This pattern suggests a trust-calibration challenge for out-of-band chat assistance: plausible explanations may appear sufficient even when users have not validated them.



\section{Discussion and Conclusion}

We presented \sysName{}, a Chrome extension that delivers in-situ assistance directly within live web interfaces through runtime DOM manipulation.
By grounding assistance in the shared structure of the web (DOM), DOMSteer demonstrates that lightweight, reversible, and application-agnostic in-situ assistance is both technically feasible and meaningfully effective. 
Through both technical evaluation and user studies, we showed that \sysName{} can reduce task completion time and improve accuracy compared with a state-of-the-art chat assistance and autonomous agent.
This work advances a vision of GUI agents that do not merely automate tasks or advise from the sidelines, but actively reconfigure the interface to support users in the moment. 



\textbf{Limitations.}
While \sysName{} is theoretically applicable to any DOM-based web interface without access to its source code, the quality of assistance depends on the structure of the underlying DOM tree. 
Interfaces that rely heavily on canvas-based rendering, heavily obfuscated class names, or deeply nested DOMs present challenges for both effective element identification and accurate modification, limiting the practical scope of deployment.
In addition, handbook construction requires 3–5 minutes of one-time initialization per interface, and the handbook size NN
N must be set empirically.
This overhead may be impractical for interfaces used infrequently or that vary substantially across sessions.

\textbf{Future Work.}
DOMSteer currently responds to explicitly stated user challenges on a per-request basis. 
Two natural extensions push toward a more proactive and systemic role. 
First, behavioral signals such as hesitation, repeated interactions with the same element, or deviation from common task paths could trigger assistance without requiring users to articulate a question. Prior work on just-in-time assistance~\cite{lam2025justintime} and interaction-trace-based intent inference~\cite{shaikh2025generalusermodel} suggests feasible paths toward this goal, and the handbook-based architecture of DOMSteer is well-positioned to support low-latency proactive delivery without invoking the LLM on every user action. 
Second, aggregating assistance patterns across users could surface recurring challenges that reflect systematic usability gaps in the interface itself (\eg, a control that consistently triggers WHAT confusion across many users). 
Such signals could inform interface-level redesign recommendations, positioning DOMSteer not only as a per-user help tool but as a lightweight, continuously updated usability auditing layer.


\bibliographystyle{ACM-Reference-Format}
\bibliography{reference}

\appendix

\section{Abridged Prompts}
\label{prompts}
We outline prompts in more detail for the in-situ assistance computational pipeline and \sysName{}.
\subsection{Assistance Generation}
\label{prompt:generation}
 
\begin{lstlisting}
Generate N={120} in-situ assistance pairs for the web
interface.
Page: {page_title} ({page_url})
Interface Knowledge: {interface_knowledge}
User challenge (optional): {user_query}
UI Elements:
  [Section] Data
  #4 select-data [button] select-data Change
  #5 [text] Cars
  [Section] Fields
  #7 [control] Cylinders
  #8 [link button] filter-field Cylinders
  #9 [link button] add-field Cylinders
  ...
 
Use the page title, URL, knowledge, and available UI elements to infer the interface's purpose, the actions it supports, and likely sources of user challenges in navigating the interface. Generate diverse and representative assistance targeting to solve the user challenges. Select the appropriate dom assistance type to resolve the challenge.  
 
{DOM_manipulation_type_design_space}
 
Common challenge types to reason about:
  - Meaning: the user does not know what a term, field, control, or output means.
  - Location: the user cannot find the right field, feature, or control.
  - Procedure: the user knows the goal but not the steps or interaction pattern.
  - Behavior: the user does not understand why the interface produced a result.
  - Direction: the user completed a step but does not know what to do next.
  - Capability: the user wants something the tool may hide, lack, or make awkward.
Cover the types that genuinely fit this UI.
 
{DOM_manipulation_type_design_space_guidelines}
 
Output schema:
{
  assistance: "[Action] [target] to [outcome]",
  whyItHelps: "Users who [need] can
    [intervention], [outcome].",
  domSubtype: one of
    1. insert.overlay_tip
    2. insert.widget
    3. insert.inline_control
    4. mutate.style
    5. mutate.representation
    6. mutate.reframe
    7. recompose.reorder
    8. recompose.group
    9. recompose.layout,
  configuration: [execution configuration of the
    DOM manipulation type],
  targets: [{ uiDescription: exact element label
    from UI element list }]
}
\end{lstlisting}
 
\noindent An example generated assistance pair:
 
\begin{lstlisting}
{
  assistance: "Insert an inline 'Search fields'
    input next to [text] Fields to locate
    [control] Production Budget quickly.",
  whyItHelps: "Users who cannot find the data field in a long data panel can type part of the field name and immediately narrow the list, reducing search effort and helping them locate the exact field faster.",
  domSubtype: "insert.inline_control",
  targets: [
    { uiDescription: "[text] Fields" }
  ],
  configuration: {
    placement: "adjacent",
    detail: {
      controlType: "search-input",
      label: "Search fields",
      placeholder: "Type a field name",
      action: {
        type: "filter-field-list",
      }
    }
  }
}
\end{lstlisting}

\subsection{Challenge Resolution}
\label{prompt:judger}
 
\begin{lstlisting}
You are evaluating whether a UI assistance
suggestion resolves a user's challenge.
 
User Need: "{user_need}"
Interface: "{interface_name}"
Generated Assistance: "{generated_assistance}"
Reference Assistance: "{annotated_assistance}"
 
Evaluate whether the generated assistance would resolve the user with their stated challenges and needs. Do not directly compare it against any reference answer or ground truth.
 
Score 0-10:
  10    = Directly and fully addresses the need
  8-9   = Addresses the need well with minor gaps
  6-7   = Mostly addresses the need but misses
          some important detail
  3-5   = Partially addresses the need
  1-2   = Tangentially related but not helpful
  0     = Does not address the need
 
Return ONLY valid JSON with the exact structure:
{
  "score": <number between 0 and 10>,
  "reasoning": "<brief explanation>"
}
\end{lstlisting}
\section{Formative Study Tasks}

\subsection{Formative Study Tasks}
\label{formative:tasks}

Participants completed a set of open-ended analysis and exploration tasks across two interfaces.

\begin{enumerate}[leftmargin=*, nosep]
\item How has the budget of movies changed over time?
\item What are the top five highest-grossing movies in the dataset? What do they have in common?
\item Within the major Action genres, identify two movies that earned the most relative to their production budgets.
\item Do movies with higher budgets tend to receive better audience ratings?
    \item Compare the distribution of audience ratings across genres. Which genre appears to be the most consistently well rated?
    \item How are audience reception and commercial success related? What example movies appear successful across multiple dimensions, such as high critical rating, high gross revenue, and/or relatively low budget?
    \item Design one neural network configuration that works across all four classification datasets. A configuration is considered to work if its decision boundary clearly separates the two classes in each dataset.
    \item Find at least one additional configuration that also works across all four datasets and differs from your first design. Explore as many alternatives as possible by varying features, network architecture (layers and neurons), and training parameters.
\end{enumerate}

\subsection{Formative Study Labeling Interface}
\label{appendix:formative-labeler}

\begin{figure}[t]
  \centering
  \includegraphics[width=\columnwidth]{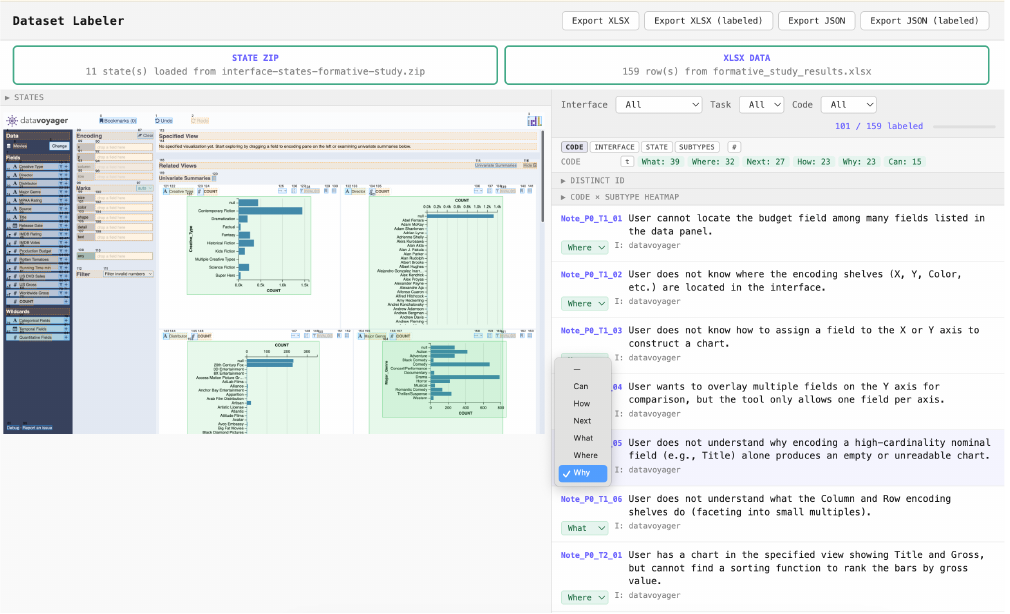}
  \caption{Interface for coding and labeling assistance-seeking instances from the formative study.}
  \label{fig:labeler}
\end{figure}

To code the 159 instances collected during the formative study, we built a custom labeling interface (\autoref{fig:labeler}) that allows researchers to review instance alongside its screen context and assign challenge-type codes.

\begin{figure}[t]
  \centering
  \includegraphics[width=0.8\columnwidth]{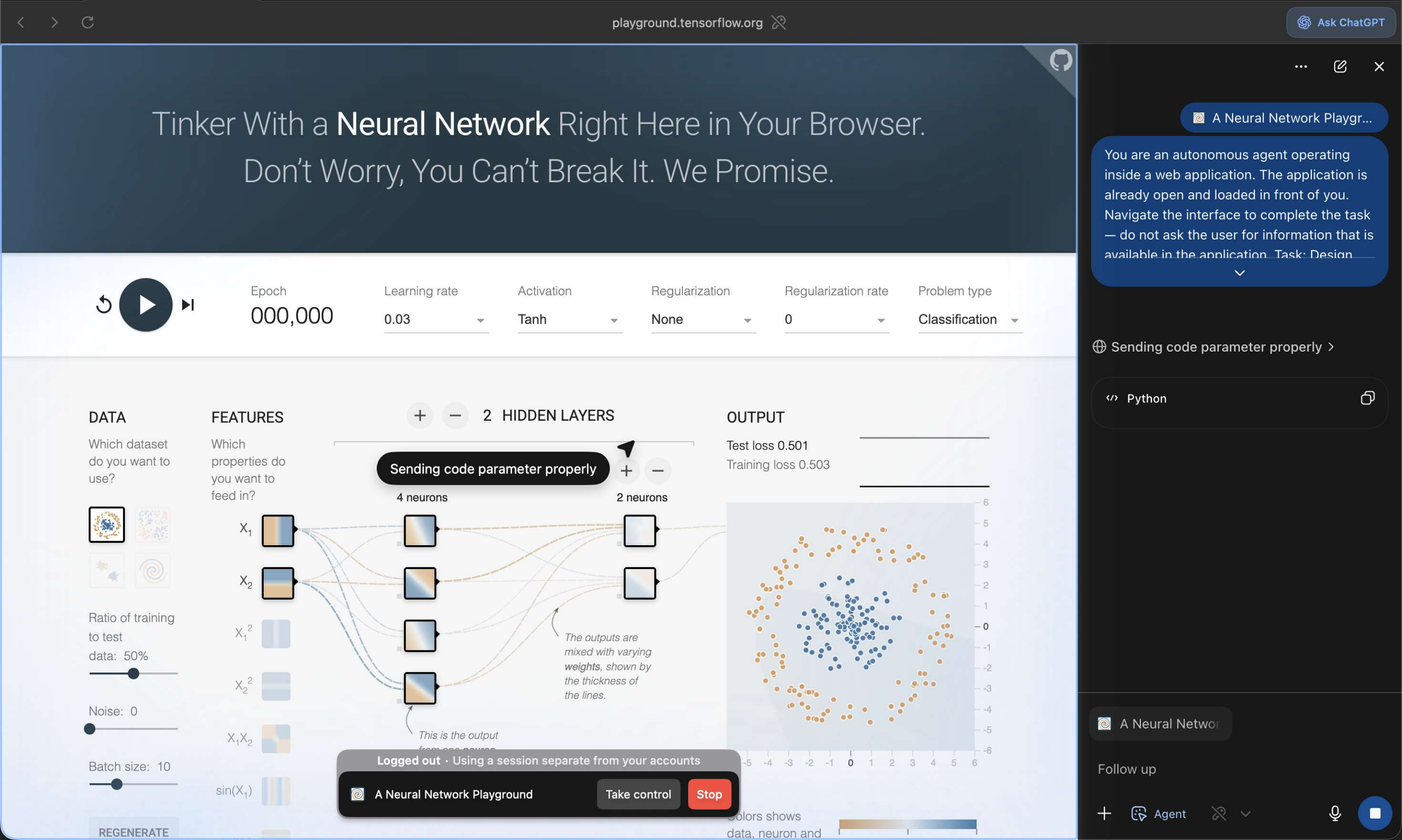}
  \caption{The autonomous assistant interface in ChatGPT Atlas.}
  \label{fig:atlas-agent-mode}
\end{figure}

\section{Evaluation: ChatGPT Atlas Baseline}
\label{appendix:atlas-baseline}
~\autoref{fig:atlas-agent-mode} shows the running of Autonomous agent condition.

\end{document}